\documentclass[preprint]{aastex}
\usepackage{graphicx}

\begin{document}

\title{Cassini UVIS Observations of the Io Plasma Torus.  \\III. 
  Observations of Temporal and Azimuthal Variability}

\author{A.~J. Steffl\altaffilmark{1} , P.~A. Delamere, and F. Bagenal}
\affil{Laboratory for Atmospheric and Space Physics, University of
  Colorado, Campus Box 392, Boulder, CO 80309-0392, USA.}
\altaffiltext{1}{Now at Southwest Research Institute, 1050 Walnut St,
  Suite 400, Boulder, CO 80302, USA}
\email{steffl@boulder.swri.edu}
\received{March 10, 2005}
\revised{July 12, 2005}
\accepted{July 29, 2005}

\shorttitle{Temporal and azimuthal variability in the Io torus}
\shortauthors{Steffl et al.}

\begin{abstract}
  In this third paper in a series presenting observations by the
  Cassini Ultraviolet Imaging Spectrometer (UVIS) of the Io plasma
  torus, we show remarkable, though subtle, spatio-temporal variations
  in torus properties. The Io torus is found to exhibit significant,
  near-sinusoidal variations in ion composition as a function of
  azimuthal position. The azimuthal variation in composition is such
  that the mixing ratio of \ion{S}{2} is strongly correlated with the
  mixing ratio of \ion{S}{3} and the equatorial electron density and
  strongly anti-correlated with the mixing ratios of both \ion{S}{4}
  and \ion{O}{2} and the equatorial electron temperature.
  Surprisingly, the azimuthal variation in ion composition is observed
  to have a period of 10.07 hours---1.5\% longer than the System III
  rotation period of Jupiter, yet 1.3\% shorter than the System IV
  period defined by \cite{Brown95}. Although the amplitude of the
  azimuthal variation of \ion{S}{3} and \ion{O}{2} remained in the
  range of 2--5\%, the amplitude of the \ion{S}{2} and \ion{S}{4}
  compositional variation ranged between 5--25\% during the UVIS
  observations. Furthermore, the amplitude of the azimuthal variations
  of \ion{S}{2} and \ion{S}{4} appears to be modulated by its location
  in System III longitude, such that when the region of maximum
  \ion{S}{2} mixing ratio (minimum \ion{S}{4} mixing ratio) is aligned
  with a System III longitude of $\sim$200$\pm15^\circ$, the amplitude
  is a factor of $\sim$4 greater than when the variation is
  anti-aligned. This behavior can explain numerous, often apparently
  contradictory, observations of variations in the properties of the Io
  plasma torus with the System III and System IV coordinate systems.
\end{abstract}

\keywords{Jupiter, Magnetosphere; Io; Ultraviolet Observations;
  Spectroscopy}


\section{Introduction}
The Io plasma torus is a dense ($\sim$2000 cm$^{-3}$) ring of
electrons and sulfur and oxygen ions trapped in Jupiter's strong
magnetic field, produced by the ionization of $\sim$1 ton per second
of neutral material from Io's extended neutral clouds. On ionization,
fresh ions tap the rotational energy of Jupiter (to which they are
coupled by the magnetic field). Much of the torus thermal energy is
radiated as intense ($\sim$10$^{12}$ W) EUV emissions. The $\sim$100
eV temperature of the torus ions indicates that they have lost more
than half of their initial pick-up energy. Electrons, on the other
hand, have very little energy at the time of ionization and gain
thermal energy from collisions with the ions (as well as through other
plasma processes) while losing energy via the EUV emissions that they
excite.

{\it In situ} measurements of the Io plasma torus from the Voyager,
Ulysses, and Galileo spacecraft and remote sensing observations from
the ground and from space-based UV telescopes have characterized the
density, temperature and composition of the plasma as well as the
basic spatial structure (see review by \cite{Thomasetal04}). However,
the azimuthal and temporal variability of the torus remains poorly
determined. Extensive measurements of torus emissions made by the
Ultraviolet Imaging Spectrograph (UVIS) on the Cassini spacecraft as
it flew past Jupiter on its way to Saturn allow us to further examine
the azimuthal structure of the plasma torus and its changes with time.

Analysis of torus spectral emissions provides estimates of plasma
composition, temperature, and density, which can then be used to
constrain models of mass and energy flow through the torus. Such
models can be used to derive plasma properties such as source
strength, source composition, and radial transport timescale
\citep{Delamere:bagenal03,Lichtenbergetal01,Schreieretal98}. Thus, one
aims to relate observations of spatial and temporal variations in
torus emissions to the underlying source, loss and transport
processes. Towards this ultimate goal, we present an analysis of UVIS
observations of the Io torus from 1 October 2000 to 14 November 2000.

\subsection{Jovian Coordinate Systems}

In order to understand the UVIS observations of azimuthal variability
and periodicity, it is useful to briefly review the various Jovian
coordinate systems (see also \cite{Dessler83} and
\cite{Higginsetal97}). Jupiter has no fixed surface features on which
measurements of its rotation period can be based. Observations of the
transits of Jovian cloud features by several late 19$^{th}$ century
astronomers, among whom were \cite{Marth1875} and \cite{Williams1896},
led to the adoption of the ``System I'' and ``System II'' rotation
periods. The System I period, based on the rate of rotation of
equatorial cloud features, was defined as 9$^h$~50$^m$~30.0034$^s$,
while the System II period, based on the more slowly rotating cloud
features at high latitudes, was defined as 9$^h$~55$^m$~40.6322$^s$
\citep{Dessler83}. The prime meridians of the associated longitude
grids for both coordinate systems were defined to be the Central
Meridian Longitude at Greenwich noon on July 14, 1897.

Attempts to derive a rotation period based on the motion of the
interior of the planet, rather than the cloud tops, met with little
success until the discovery of decametric (DAM; $\sim$20 MHz) radio
emissions from Jupiter by \cite{Burke:franklin55}. From this radio
emission, \cite{Shain56} obtained the first measurement of the Jovian
rotation period based on the rotation of the magnetosphere. Subsequent
observations improved the accuracy of the rotation period derived from
radio emissions, and a period of 9$^h$~55$^m$~29.37$^s$---the weighted
average of several radio observations---was defined as the ``System
III'' (1957) rotation period \citep{Burkeetal62}. With further
observations, it gradually became clear that this period was in need
of a slight revision. \cite{Riddle:warwick76} reported the weighted
average of radio observations obtained since the System III (1957)
period was defined; their published value of 9$^h$~55$^m$~29.71$^s$
became known as System III (1965) and was adopted by the International
Astronomical Union (IAU) as the standard rotation period of Jupiter
\citep{Seidelman:devine77}. Recently, \cite{Higginsetal97} reported
that the System III (1965) rotation period should be further revised
to 9$^h$~55$^m$~29.6854$^s$, based on 35 years of radio observations
of Jupiter. This difference results in a shift of $\sim$1$^{\circ}$ in
longitude every four years relative to System III (1965). Since this
shift in longitude is far less than can be measured by UVIS during the
Jupiter encounter and since the proposed revision to the System III
period has not yet been uniformly adopted \citep{Russelletal01}, all
subsequent references to ``System III'', corotation, or rotation with
the magnetic field, will refer to the IAU accepted Jovian rotation
period: System III (1965).

\subsection{Variations with System III Longitude}

For the purposes of modeling, the Io plasma torus is often assumed to
be azimuthally symmetric. However there have been numerous
observations of the Io torus that suggest that the torus exhibits
significant variation with System III longitude. Here, we present a
brief review of some of these observations. An additional discussion
of observations of longitudinal asymmetries can be found in
\cite{Thomas93b}.

Some of the earliest observations of the Io plasma torus found that
the brightness of the [\ion{S}{2}] 6716\AA/6731\AA\ doublet was
correlated with System III longitude. \cite{Traugeretal80} observed
this [\ion{S}{2}] doublet using the 5~m Hale telescope on the nights
of 7--11 October 1976. They found that a region extending 90$^{\circ}$
in longitude and centered on $\lambda_{III}=280^{\circ}$ was
consistently fainter than the rest of the torus. The [\ion{S}{2}]
brightness peaked at $\lambda_{III} \approx 180^{\circ}$, although
given the relatively large scatter in the data, this value is poorly
constrained.

Using the 2.2~m telescope of the Mauna Kea observatory,
\cite{Pilcher:morgan80} observed the [\ion{S}{2}] 6716\AA/6731\AA\
doublet over a three month interval that began in December 1977. The
brightness of the [\ion{S}{2}] doublet was found to vary with
longitude by as much as a factor of 4. The peak brightness was
observed in the longitude range of 160$^{\circ} < \lambda_{III} <
340^{\circ}$. At other times, \cite{Pilcher:morgan80} found the
[\ion{S}{2}] brightness to be more azimuthally uniform, with the
transition between these two states taking approximately two weeks.

\cite{Trafton80} reported a similar azimuthal variation in the
brightness of the [\ion{S}{2}] 6716\AA/6731\AA\ doublet in
widely-spaced observations between 19 January 1976 and 19 June 1979
using the McDonald Observatory's 2.7~m telescope. The brightness was
found to vary by about a factor of 5, with a peak located at
$\lambda_{III}$=260$^{\circ}$.

Extensive observations of the Io plasma torus were made by the
ultraviolet spectrometers (UVS) aboard the Voyager 1 and Voyager 2
spacecraft \citep{Broadfootetal77,Broadfootetal81}. The initial search
for variations in the UV brightness of the torus with System III
longitude focused on the pre-encounter period of the Voyager 2
spacecraft, which took place from days 116--144 of 1979
\citep{Sandel:broadfoot82a}. During the period of day 121/16:20 UT to
day 123/14:00 UT, a weak ($<$10\%) azimuthal variation in the
brightness of the \ion{S}{3} 685\AA\ feature was observed. This
variation had a peak brightness located in the range of $330^{\circ} <
\lambda_{III} < 40^{\circ}$ and a minimum brightness in the range of
$140^{\circ} < \lambda_{III} <200^{\circ}$ and was seen only in the
dusk ansa of the torus. Roughly two days later, during the period of
day 123/18:00 UT to day 125/12:00 UT, a stronger azimuthal variation
was seen in both the dawn and dusk ansae. However, the phase of the
variation had shifted by $\sim$60$^{\circ}$ in longitude, such that
the peak in brightness was located between $40^{\circ} <\lambda_{III}
< 100^{\circ}$ and the minimum between $180^{\circ} <\lambda_{III} <
240^{\circ}$.

A short-term variation with System III longitude similar to that
reported by \cite{Sandel:broadfoot82a} has been found in spectra from
the Voyager 1 UVS \citep{Herbert:sandel00}. After analyzing 47 hours
of Voyager 1 UVS spectra of the Io torus, \cite{Herbert:sandel00}
found that both the electron density and electron temperature vary
with System III longitude. The electron density variation had an
amplitude of about 12\% with a peak near $\lambda_{III}$=150$^{\circ}$
and a minimum near $\lambda_{III}$=320$^{\circ}$. the electron
temperature variation had an amplitude of about 7\% with a a peak near
$\lambda_{III}$=270$^{\circ}$ and a minimum near
$\lambda_{III}$=80$^{\circ}$.

Although \cite{Sandel:broadfoot82a} found evidence for short-term
azimuthal variations in the brightness of the \ion{S}{3} 685\AA\
feature, they reported no significant long-lived variation of torus
brightness with System III longitude during the 44-day Voyager 2
pre-encounter period. However, such a long-lived System III variation
was found by \cite{Sandel:dessler88}, who used a Lomb-Scargle
periodogram analysis \citep{Lomb76,Scargle82, Horne:baliunas86} to
search for periodicities in the Voyager 2 pre-encounter data. Using
similar analysis techniques (Lomb-Scargle periodograms)
\cite{Woodwardetal94} and \cite{Brown95} also discovered System III
periodicities in their observations of the torus [\ion{S}{2}] 6731\AA\
emission.

Imaging observations of the torus in 1981 by \cite{Pilcheretal85}
showed that the brightness of the [\ion{S}{2}] 6731\AA\ line varied by
a factor of 6 as a function of System III longitude. The peak
brightness was found to be at $\lambda_{III} \approx 170^{\circ}$,
although a secondary peak at $\lambda_{III} \approx 280^{\circ}$ was
also evident.

Also in 1981, \cite{Morgan85} was able to simultaneously image the
[\ion{S}{2}] 6716\AA/6731\AA\ doublet, the [\ion{S}{2}]
4069\AA/4076\AA\ doublet, and the [\ion{O}{2}] 3726\AA/3729\AA\
doublet, using the Mauna Kea Observatory 2.2~m telescope. On observing
runs that took place from 14--17 February 1981 and 20--23 March 1981,
\cite{Morgan85} found the brightness of both [\ion{S}{2}] doublets
varied with System III longitude, with a peak at $\lambda_{III}
\approx 180^{\circ}$. However, no correlation of the brightness of the
[\ion{O}{2}] doublet with System III longitude was apparent.
\cite{Brown:shemansky82} made spectroscopic observations of the Io
torus [\ion{S}{2}] 6716\AA/6731\AA\ doublet on 23--24 February
1981---six days after the observations of \cite{Morgan85}---and found
no obvious correlation of [\ion{S}{2}] brightness with System III
longitude.

Additional imaging of the Io plasma torus at the [\ion{S}{2}] 6731\AA\
emission was conducted by \cite{Schneider:trauger95} over six nights
from 31 January 1991 to 6 February 1991. They found the longitudes
$150^{\circ} < \lambda_{III} < 210^{\circ}$ to be consistently
$\sim$3--4 times brighter than the longitudes $0^{\circ} <
\lambda_{III} < 170^{\circ}$. More detailed examination revealed that
the variation of brightness with longitude was weakest on 31 January
1991 with a poorly-constrained maximum near $\sim$120$^{\circ}$. Three
nights later, the variation with longitude was significantly stronger,
and the peak had shifted to a longitude of $\sim$170$^{\circ}$.
Finally, on the last night of observation (5 February 1991), the
amplitude of the longitudinal variation remained relatively large, and
the peak had shifted further to a longitude of $\sim$210$^{\circ}$.
\cite{Schneider:trauger95} interpreted the shift in phase of
$\sim$18$^{\circ}$/day as evidence for a possible 2.1\% subcorotation
(relative to rigid corotation) of a torus feature.  Additionally,
\cite{Schneider:trauger95} proposed that the modulation of the
amplitude of the longitudinal variation might explain why numerous
previous observations had detected an enhanced brightness of
\ion{S}{2} in the ``active sector'' (a sector spanning roughly
90$^{\circ}$ in longitude centered around $\lambda_{III} \approx
180^{\circ}$): the amplitude of the variation is greatest when the
peak lies within this region and is diminished when it lies outside.

Spectra of the Io torus obtained on 10--11 February 1992 showed that
the brightness of the [\ion{S}{2}] 6716\AA/6731\AA\ doublet and the
\ion{S}{3} 6312\AA\ line were correlated. These spectral features
peaked in brightness at a System III longitude of $\approx
180^{\circ}$ \citep{Raueretal93}.

In contrast to observations of the \ion{S}{3} 685\AA\ feature by the
Voyager 1 and Voyager 2 UVS \citep{Sandel:broadfoot82a,
  Herbert:sandel00}, \cite{Gladstone:hall98} found no correlation
between the brightness of torus emission between 70--760\AA\ and
System III longitude.

Emission from the [\ion{S}{4}] 10.51 $\mu$m line was discovered in
observations of the Io plasma torus on 25 May 1997 using the Infrared
Space Observatory \citep{Lichtenbergetal01}.  The brightness of this
feature was found to vary by $\sim$20\% with System III longitude with
a poorly-constrained peak near $\lambda_{III} \approx 120^{\circ}$.

\subsection{Subcorotating Torus Phenomena and ``System IV''}

In the 29 years since its discovery by \cite{Kupoetal76}, there have
been numerous observations of phenomena occurring in the Io plasma
torus having a period longer than the System III rotation period.
There have also been several direct measurements of the torus plasma
lagging corotation with System III. To place our results into proper
context, we present a brief review of these observations.

\subsubsection{Radio Emissions}
The first indications that plasma in the Io torus might not be rigidly
corotating with Jupiter's magnetic field came from the Planetary Radio
Astronomy (PRA) experiments aboard the two Voyager spacecraft. The
Jovian narrow-band kilometric radiation (nKOM), first described by
\cite{Kaiser:desch80}, is emitted from source regions lying in the
outer Io plasma torus at radial distances of $\sim$8--9~R$_J$.
\cite{Kaiser:desch80} found that the rotation period of Jovian
narrow-band kilometric radiation (nKOM) source regions was 3.3\%
slower than the System III rotation period during the Voyager 1
encounter and 5.5\% slower during the Voyager 2 encounter.

The initial analysis by \cite{Kaiser:desch80} of the Jovian nKOM
emissions observed by the Voyager PRA experiments covered only a
relatively short period ($\sim$45 Jovian rotations) during each
spacecraft encounter. A statistical analysis of all detections of nKOM
by both Voyager PRA experiments between 14 January 1979 and 31
December 1979 (the period when the spacecraft were within 900~R$_J$
of Jupiter) found that the rotation period of the nKOM sources was not
constant \citep{Daigne:leblanc86}. Rather, the rotation periods for
individual nKOM sources varied between 0--8\% longer than the System
III rotation period with average values of 3.2\% and 2.7\% for Voyager
1 and Voyager 2, respectively. The large range of values reflects
intrinsic variability in the rotation period of the nKOM sources
rather than errors in measurement, which are estimated to be
$\sim$1\%. Although the rotation period of individual nKOM sources
generally lagged the System III rotation period, the probability of
observing nKOM emission was found to be significantly greater when the
spacecraft were at System III longitudes of 40$^{\circ}$ and
300$^{\circ}$.

The order of magnitude greater sensitivity and direction-finding
capabilities of the Unified Radio and Plasma Wave instrument (URAP) on
the Ulysses spacecraft allowed the detection of six distinct nKOM
source regions during the Ulysses encounter with Jupiter in February,
1992 \citep{Reineretal93}.  These source regions were found to lie at
radial distances of 7.0--10.0 R$_J$ and to have a rotation periods
ranging from 3.0--8.6\% greater than the System III period (again the
range in values represents the variability of the individual nKOM
sources, rather than measurement uncertainty). In addition, to the
subcorotation period of nKOM source regions, URAP also detected a new
component of the Jovian hectometer radiation (HOM) that recurs with a
period 2--4\% longer than the System III rotation period
\citep{Kaiseretal96}. Reexamination of Voyager 1 and Voyager 2 PRA
data found similar results. Based partly on the spectroscopic
observations of \cite{Brown95}, which were concurrent with the Ulysses
encounter, \cite{Kaiseretal96} conclude that the new HOM component is
the result of an HOM source region in the high-latitude regions of
Jupiter being periodically blocked by a high-density region in the Io
torus.

\subsubsection{{\it In Situ} Plasma Measurements}

{\it In situ} measurements of the bulk rotation velocity of the torus
plasma have been made by the Plasma Science (PLS) instruments aboard
the Voyager 1 and Galileo spacecraft and the URAP instrument aboard
the Ulysses spacecraft. The Voyager 1 PLS found that the torus plasma
was within a couple percent of rigid corotation inside of 5.7~R$_J$,
but between 5.9--10 R$_J$, deviations from corotation of up to 10\%
could not be ruled out \citep{Bagenal85}. The Ulysses URAP instrument
measured two components of the dc electric field during its fly-through
of the outer Io torus, and from this derived the flow speed of torus
plasma \citep{Kelloggetal93}. URAP found the plasma flow speeds to be
generally close to corotation but with significant deviations having
an rms value of 5.3 km/s. Finally, on five passes through the Io
torus, the Galileo PLS observed that the bulk plasma flow lagged the
corotation velocity by 2--10 km/s, with an average deviation of
$\sim$2--3 km/s \citep{Frank:paterson01a}.

\subsubsection{Spectroscopic Measurements}
\label{sys4_spectroscopic}

The first direct observation of a corotational lag in the Io torus
plasma came from analysis of the Doppler shift of the
[\ion{S}{2}]~6716~\AA/6731~\AA\ doublet \citep{Brown83}. Using
observations from two nights in February, 1981 and three nights in
April, 1981, \cite{Brown83} found that the radial velocity of
\ion{S}{2} deviated from rigid corotation by 6\%$\pm$4\%, where the
$\pm$4\% represents the variability of the derived corotation lag,
rather than the measurement uncertainty.

Observations of [\ion{S}{3}]~9531~\AA\ emitted from the dusk side of
the torus between 12 April 1982 and 30 April 1982 found that the torus
brightness was not correlated with the System III rotation period, but
rather with a period of 10.2$\pm$0.1 hours, 2.8\% longer than the
System III period \citep{Roesleretal84}. These observations were
obtained between 12 April 1982 and 30 April 1982 using a scanning
Fabry-Perot spectrometer that had a field of view 2~R$_J$ in diameter
centered at a radial distance of 6~R$_J$. This was the first detection
of a long-lived (roughly 43 rotations of Jupiter) periodic phenomenon
in the Io torus at a period other than System III. A reanalysis of
this data by \cite{Woodwardetal94} confirmed the existence of a
10.20~hour periodicity in the data and found a statistically
significant secondary periodicity at the System III rotation period.
Additional ground-based observations of [\ion{S}{3}]~9531~\AA and
[\ion{S}{2}]~6731~\AA\ emission from the Io torus in March and April,
1981 by \cite{Pilcher:morgan85} and \cite{Pilcheretal85} were
interpreted as requiring a torus rotation period a few percent longer
than System III, consistent with \cite{Roesleretal84}.

The reported subcorotation of the nKOM source regions
\citep{Kaiser:desch80}, the 10.2-hour periodicity in the brightness of
[\ion{S}{3}]~9531\AA\ \citep{Roesleretal84}, and analysis of Voyager
Ultraviolet Spectrometer (UVS) data \citep{Sandel83}, led
\cite{Dessler85} to propose a new Jovian coordinate system known as
``System IV'' that rotates 3.1\% slower than System III. The proposed
System IV coordinate system was further refined by
\cite{Sandel:dessler88}. Using a Lomb-Scargle periodogram analysis of
the brightness of the Voyager 2 UVS 685~\AA\ feature, (a feature
dominated by three multiplets of \ion{S}{3} though also containing
emissions from \ion{S}{4} and \ion{O}{3}), \cite{Sandel:dessler88}
found evidence for periodicity at both the System III period, and at a
period of 10.224 hours, 3.0\% longer than System III.  In addition,
\cite{Sandel:dessler88} noted that the azimuthal variation in
brightness was greatest when these two periods were aligned. The
observed period of 10.224 was used to define the System IV (1979)
coordinate system. The prime meridian of the System IV (1979)
coordinate system was defined such that the peak of the 685~\AA\
emissions occurred near $\lambda_{IV}$=180$^{\circ}$. Recent analysis
of 47 hours of Voyager 1 UVS spectra of the Io torus by
\cite{Herbert:sandel00} found both electron density and electron
temperature to be organized in System IV longitude. However, the
uncertainty in these quantities is larger than the observed System IV
modulation.

The first observational program designed specifically to look for
periodicities in the Io torus was undertaken in 1988
\citep{Woodwardetal94}. \cite{Woodwardetal94} observed emission of
[\ion{S}{2}]~6731\AA\ from the Io torus over a 35 day period using a
Fabry-Perot spectrometer similar to that used by \cite{Roesleretal84}.
After careful analysis of their data using weighted Lomb-Scargle
periodograms, they found periodicity in the torus [\ion{S}{2}]
intensity at 10.14$\pm$0.03 hours---a period intermediate of System
III and System IV---and at 9.95 hours, consistent with the System III
period.

In an effort to address the apparent inconsistencies between the
previous spectroscopic measurements, \cite{Brown94a} observed the
[\ion{S}{2}] doublet at 6717\AA\ and 6731\AA\ over a six month period
in 1992 using a long-slit echelle spectrograph. To date, this remains
the longest time baseline of torus observations, and it thus provides
the most accurate measurement of the periodicities in the torus. Using
the now ubiquitous Lomb-Scargle periodogram analysis, \cite{Brown95}
found significant periodicity in the torus at both the System III
period and a period of 10.214$\pm$0.006 hours---2.91\% longer than the
System III period. The latter period was found to remain constant
between the radial distances of 5.875 and 6.750 R$_J$ and provides the
basis for a minor revision to the System IV (1979) coordinate system
known as System IV (1992). Subsequent references to ``System IV'' will
refer to the System IV (1992) period defined by \cite{Brown95}.

During the observing period, the variation of [\ion{S}{2}] line
brightness with System IV longitude underwent a sudden phase shift of
$\sim$100$^{\circ}$. The sudden shift in phase resulted in a spurious
weak peak in the dawn ansa periodogram at a period of 10.16
hours---quite similar to the value reported by \cite{Woodwardetal94}.
By subdividing the data into two groups (before and after the sudden
phase shift) the peak in the dawn ansa periodogram becomes
10.217$\pm$0.010 hours, consistent with the value of 10.214$\pm$0.006
hours obtained from the dusk ansa. In light of the discovery that the
phase of System IV variations can shift rapidly, \cite{Woodwardetal97}
reanalyzed their 1988 data and found a similar phase shift was
responsible for the reported periodicity of 10.14$\pm$0.03 hours. By
subdividing their data into two groups they found that the primary
periodicity in the data was, in fact, at 10.2 hours, consistent with
the System IV period.

Direct measurements of the radial velocity of the torus plasma as a
function of radial distance were obtained by measuring the Doppler
shift in the sum of all 222 spectra obtained in 1992 \citep{Brown94b}.
The torus plasma was found to lag rigid corotation with System III,
with the amount of corotational lag being a strong function of radial
distance. The corotational lag reached a maximum deviation of $\sim$4
km/s in the range of 6--6.5~R$_J$. Between 6--7~R$_J$, the torus lags
corotation by an average of 2\%. This measurement, coupled with the
observation that the System IV period remained constant between 5.875
and 6.75 R$_J$, led \cite{Brown95} to conclude that the System IV
periodicity cannot be caused by plasma lagging corotation.

The radial velocity profile of the Io torus was measured again in
October 1999 using the 3.5~m European Southern Observatory's New
technology Telescope (NTT) \citep{Thomasetal01}. The radial velocities
derived from \ion{S}{2} and \ion{S}{3} emission lines were in good
agreement with the range in velocities measured by \cite{Brown83} and
\cite{Brown94b}, however the larger collecting area of the NTT
telescope enabled \cite{Thomasetal01} to place much smaller relative
error bars on their radial velocity profiles. The radial velocity
measurements of \cite{Thomasetal01} represent a snapshot of the Io
torus radial velocity profile (they were derived from a single
integration) whereas the radial velocity profile of \cite{Brown94b}
represents the average profile over a six-month period.

Finally, a multi-year campaign to determine the long-term variability
of torus [\ion{S}{2}] 6731\AA\ and 6716\AA\ emissions has been carried
out by \cite{Nozawaetal04}. Using small telescopes (diameters of 28~cm
and 35~cm), \cite{Nozawaetal04} obtained data in four observing
seasons between 1997 and 2000. Using Lomb-Scargle periodogram
analysis, they found periodicities of 10.18$\pm$0.06 hours in 1998,
10.29$\pm$0.14 hours in 1999, and 10.14$\pm$0.11 hours in 2000, all
within measurement uncertainty of the 10.214 hour System IV period.
Data from the 2000 observing season were acquired between 15 December
2000 and 5 January 2001, concurrent with the Cassini spacecraft's
closest approach with Jupiter, but more than 30 days after the data
presented in this paper were acquired.

\section{Observations and Data Analysis
  \label{observation_section}}
The data used in this paper were obtained by the Cassini spacecraft's
Ultraviolet Imaging Spectrograph (UVIS) \citep{Espositoetal04} between
1 October 2000 and 15 November 2000 (DOY 275--320) during the inbound
leg of Cassini's Jupiter flyby.  During this period, while the
spacecraft was between 1100~R$_J$ and 600~R$_J$ from the planet
(1~R$_J$=71,492 km), 1904 spectrally-dispersed images of the Io torus,
in its entirety, were acquired. All of these spectral images have an
integration time of 1000 seconds. The duty cycle for UVIS consisted of
six 20-hour blocks. During blocks 1, 2, 5, and 6 UVIS observed the Io
torus for 9 consecutive hours followed by 11 hours of downlink and
observations of other targets. Blocks 3 and 4 consisted of 28 hours of
torus observation followed by 12 hours of downlink and other
observations. This cycle was repeated nine times. Additional
information about this dataset, including examples of the observing
geometry, images of the raw and processed data, and descriptions of
the data reduction and calibration procedures used, can be found in
\cite{Steffletal04a}.

Since the encounter distances were so large, the spatial resolution of
this dataset is relatively coarse (0.6--1.1~R$_J$ per detector row).
We therefore limited our analysis to spectra from the ansa region on
both sides (dawn and dusk) of the torus. The ansa region was defined
as the part of the torus subtended by the brightest row on the
detector, plus the two neighboring rows. Spectra contained in these
three rows were averaged together to obtain the ansa spectrum. The
decreasing distance of the Cassini spacecraft to Jupiter during the
observation period meant that the range of projected radial distances
in the Io torus from which the ansa spectra were extracted went from
4.5--8.0~R$_J$ on 1 October 2000 to 5.2--7.0~R$_J$ on 15 November
2000.

The Io torus spectral model described in \cite{Steffletal04b} was used
to derive the ion composition, electron temperature, and electron
column density from the spectra extracted from each ansa of the torus.
Spectra from the dawn and dusk ansae of the torus are fit
independently of each other and yield statistically identical results.
For clarity of presentation, all figures (with the exception of
Fig.~\ref{ne_te_vs_sys3}) show results from only one of the torus
ansae.

Since the long axis of the UVIS entrance slit was oriented parallel to
the Jovian equator, information about the latitudinal distribution of
the Io torus is convolved with spectral information along the
dispersion direction of the detector.  To separate theses effects, we
assume a Gaussian scale height for the torus plasma, the value of
which is a parameter fit by the model.  Additionally, we assume that
the scale heights for all ion species present in the torus are equal.
Although this last assumption is somewhat unphysical, given the
relatively coarse spatial resolution of the UVIS dataset, it has no
significant effect on our results.

\section{Results}
\subsection{Temporal Variation in Torus Composition}
\begin{figure}[p]
  \includegraphics[scale=.75]{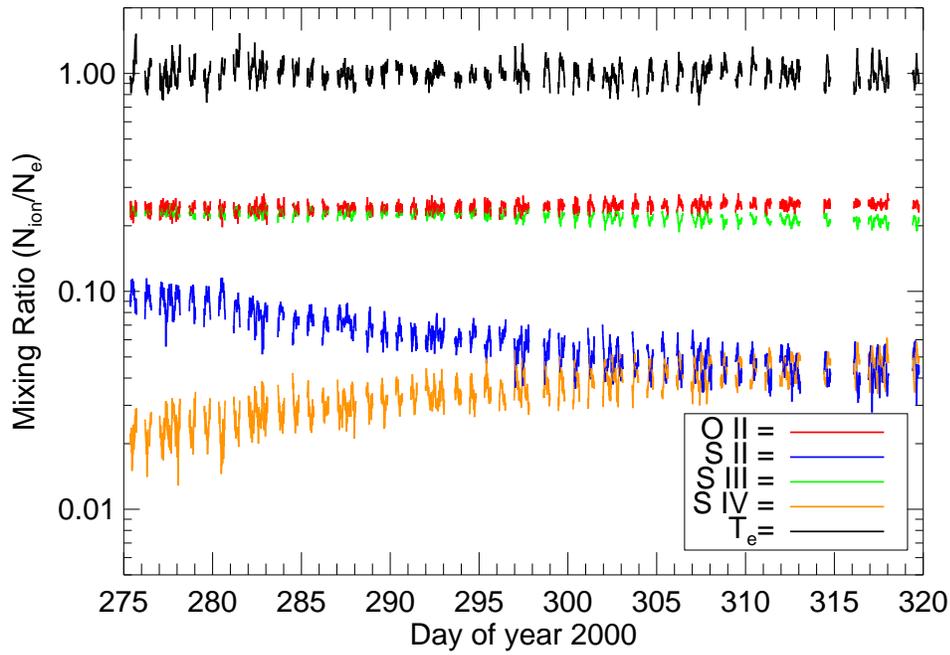}
  \caption[]
  {Ion mixing ratios (ion column density divided by electron column
    density) and electron temperature versus time, as derived from the
    dusk ansa of the torus. Owing to uncertainty in the absolute
    calibration of the UVIS Extreme Ultraviolet (EUV) channel below
    800~\AA, the electron temperature is presented in relative units.
    Results from the dawn ansa are similar. \label{p3_mix_vs_time}}
\end{figure}

Figure~\ref{p3_mix_vs_time} shows how the mixing ratios (ion density
divided by electron density) of four ion species in the torus:
\ion{S}{2}, \ion{S}{3}, \ion{S}{4}, and \ion{O}{2}, and electron
temperature vary with time during the observation period.  The most
obvious long-term change is that the mixing ratio of \ion{S}{2} falls
from 0.10 to 0.05 over a 45-day timescale, while the mixing ratio of
\ion{S}{4} increases from 0.02 to 0.05 over the same period. The
mixing ratios of \ion{O}{2} and \ion{S}{3}, the two dominant ion
species in the torus, remain relatively constant. The temporal changes
in the composition of the torus plasma coupled with observations by
the Galileo Dust Detector System of a four-orders-of-magnitude
increase in the amount of dust emitted from Io \citep{Kruegeretal03}
led \cite{Delamereetal04} to propose a factor of 3--4 increase in the
amount of neutral material available to the torus on, or around, 4
September 2000. Torus chemistry models including such an increase in
the neutral source rate (along with a corresponding increase in the
amount of hot electrons in the torus) can closely match the observed
changes in plasma composition with time.

\subsection{Azimuthal Variations in Torus Composition
  \label{azimuthal_variations_subsection}}

\begin{figure}[tbp]
  \includegraphics[scale=.75]{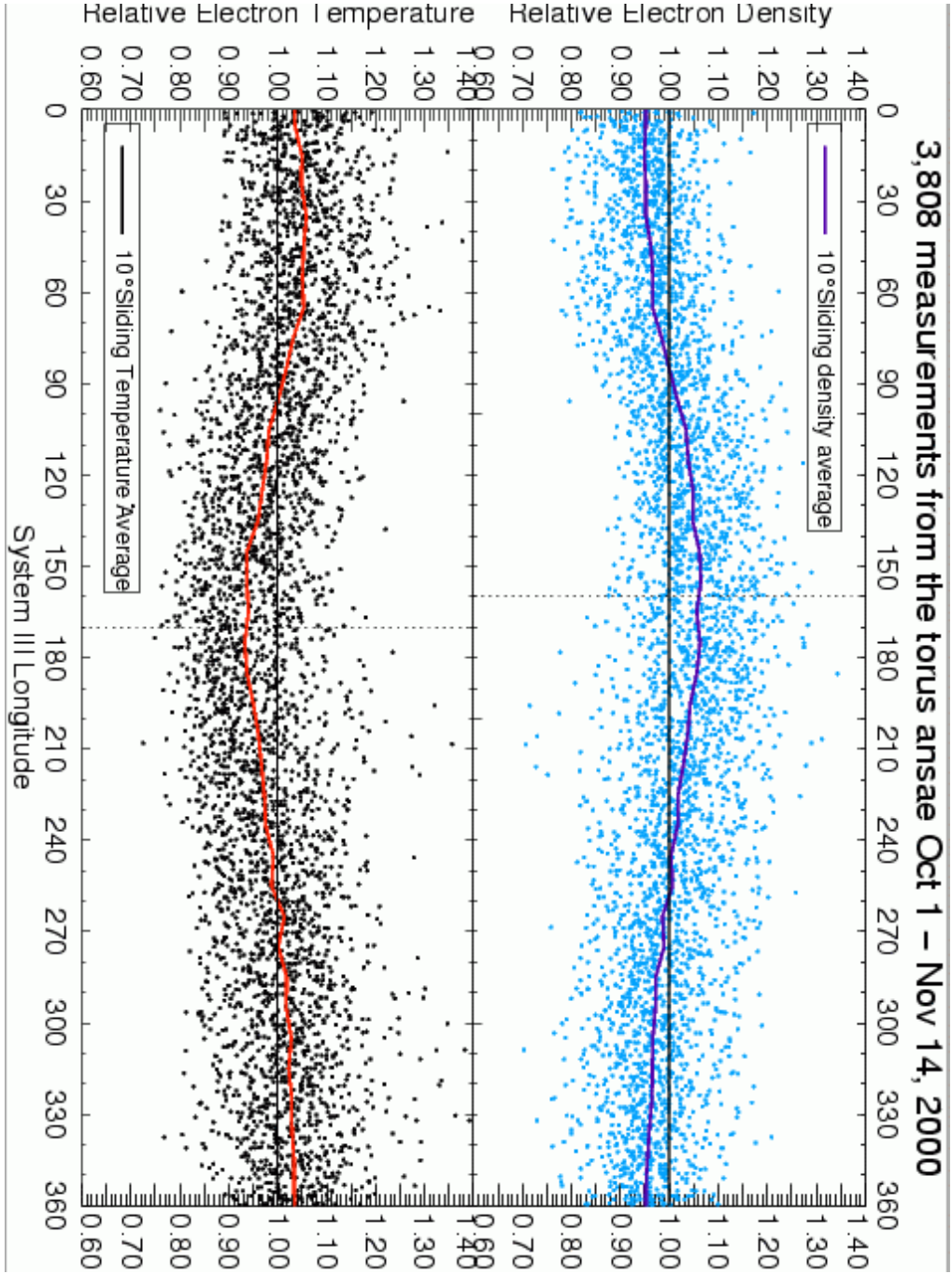}
  \caption[]
  {Relative torus electron density and electron temperature versus
    System III longitude from both dawn and dusk ansae. Values have
    been normalized to the azimuthally-averaged value at the time each
    observation was made. The solid lines represent the average of the
    data in 10$^{\circ}$ longitude bins. Both electron density and
    electron temperature show a long-term correlation with System III
    longitude of $\sim$5\%.  Although the scatter of the individual
    data points is considerable, on average, the electron density is
    anti-correlated with the electron temperature.
    \label{ne_te_vs_sys3}}
\end{figure}

Over the 45-day inbound staring period, the long-term variations of
torus parameters with System III longitude are relatively small. The
relative variation of the EUV luminosity of the torus ansae with
System III longitude is only about 5\%, with a maximum near
$\lambda_{III}$=120$^{\circ}$ and a secondary peak near
$\lambda_{III}$=270$^{\circ}$ \citep{Steffletal04a}). The relative
variations of electron density and electron temperature with System
III longitude are shown in Fig.~\ref{ne_te_vs_sys3}. Like the EUV
luminosity, both electron density and electron temperature show
long-tern variations of only $\sim$5\%. In contrast, however, the
variations of both electron density and electron temperature show a
single, clearly defined peak with a single, clearly defined minimum.
Although there is a large amount of scatter in the individual data
points, the average variation in electron density is clearly
anti-correlated with the average variation in electron temperature:
the electron density has a maximum value near
$\lambda_{III}$=160$^{\circ}$, while the electron temperature has a
minimum value near $\lambda_{III}$=170$^{\circ}$.

\begin{figure}[tbp]
  \includegraphics[scale=.75]{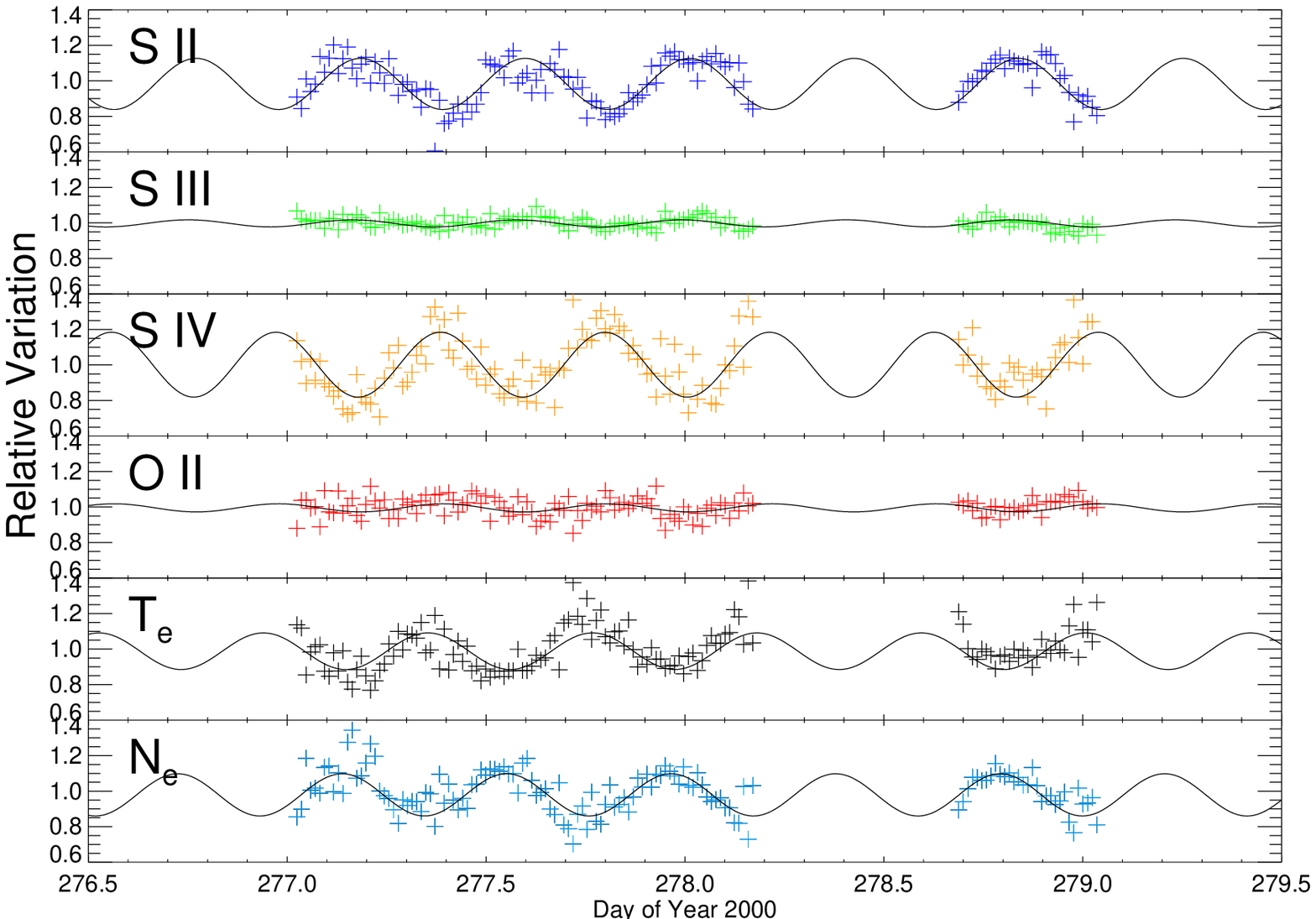}
  \caption[]
  {Relative ion mixing ratios, electron temperature, and electron
    column density for a typical 3-day period obtained from the dusk
    ansa. Values are normalized to the average value over the 3-day
    period. The best-fit sinusoids for this period are overplotted.
    Note the strong anti-correlation of \ion{S}{2} with \ion{S}{4} and
    equatorial electron temperature with equatorial electron column
    density.\label{mix_sinusoids_vs_time}}
\end{figure}

Although the torus exhibits relatively small variations with System
III longitude on long timescales, azimuthal variations of up to 25\%
are observed on timescales of a few days. This can be seen in the
high-frequency component of the curves in Fig.~\ref{p3_mix_vs_time}.
All four ion mixing ratios, as well as the relative electron
temperature and column density, exhibit near-sinusoidal variations
with a period close to that of the 9.925-hour System III (1965)
rotation period of Jupiter. This can be more readily seen in
Fig.~\ref{mix_sinusoids_vs_time}, which shows the mixing ratios of
\ion{S}{2}, \ion{S}{3}, \ion{S}{4}, and \ion{O}{2}; the electron
temperature, and the electron column density as derived from the dusk
ansa of the torus over a typical three day period (DOY 276.5--279.5).
Results from the dawn ansa are virtually identical to those from the
dusk ansa presented here, but the phase is shifted by 180$^{\circ}$.
For ease of comparison, we present these quantities normalized to
their average value over the three day period. The overplotted solid
curves are best-fit sinusoids to the data. These sinusoids have a
period equal to the System III rotation period of Jupiter.

The mixing ratios of \ion{S}{2} and \ion{S}{4}, and the electron
temperature and column density show variations of roughly 25\% over
this three day period, while the mixing ratios of \ion{S}{3} and
\ion{O}{2} show variations near the 5\% level. In addition, the
variations of \ion{S}{2}, \ion{S}{3}, and electron column density are
close to being in phase with each other, while at the same time, they
are nearly 180$^{\circ}$ out of phase with the variations in
\ion{S}{4}, \ion{O}{2}, and electron temperature. The strong
anti-correlation between the mixing ratios of \ion{S}{2} and
\ion{S}{4} can also be seen in the brightnesses of individual spectral
features. Figure~\ref{sii_vs_siv} shows the integrated luminosity from
the dusk half of the torus for two pairs of \ion{S}{2} and \ion{S}{4}
lines. Since the spectral features in each pair are close to each
other in wavelength, they have roughly the same excitation energies
and therefore are unaffected by variations in electron temperature.
Additionally, the confidence interval analysis presented in Fig. 8 of
\cite{Steffletal04b} shows that the \ion{S}{2} and \ion{S}{4} mixing
ratios derived from the spectral fitting model are only very weakly
correlated and therefore, the strong anti-correlation of these
quantities cannot be merely an artifact of the spectral fitting model.

Given that the only significant production mechanism for \ion{S}{4} is
the electron impact ionization of \ion{S}{3} (which becomes more
efficient at higher temperatures for values typical of the Io torus):

\begin{figure}[tbp]
  \includegraphics[scale=.75]{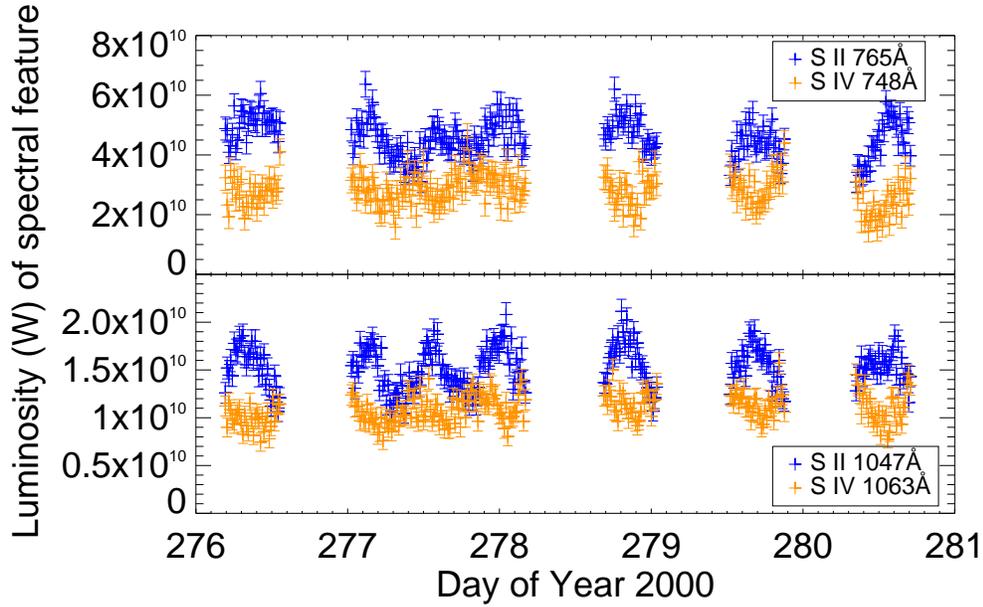}
  \caption[]
  {Integrated luminosity for two pairs of closely-spaced \ion{S}{2}
    and \ion{S}{4} spectral features from the dusk half of the torus.
    Although the error bars for an individual spectral feature are
    significantly larger than the error bars for the mixing ratios
    derived from the simultaneous fit to all spectral features(i.e.
    Fig~\ref{mix_sinusoids_vs_time}), the anti-correlation of
    \ion{S}{2} and \ion{S}{4} is readily apparent.\label{sii_vs_siv}}
\end{figure}

\begin{equation}
  \label{s2+_ionization_eqn}
  S^{2+}+e^- \rightarrow S^{3+}+2e^-
\end{equation}

\noindent it is not surprising that the \ion{S}{4} mixing ratio is
positively correlated with electron temperature. For \ion{S}{2},
electron impact ionization is both a source and a loss process:

\begin{eqnarray}
  \label{s_ionization_eqn} S+e^- \rightarrow S^{+}+2e^- \\
  \label{s+_ionization_eqn} S^++e^- \rightarrow S^{2+}+2e^-
\end{eqnarray}

\noindent As the electron temperature increases the loss rate of
\ion{S}{2} via Eq.~\ref{s+_ionization_eqn} more rapidly than the
production rate via Eq.~\ref{s_ionization_eqn}, \ion{S}{2} should show
an anti-correlation with electron temperature, which is what is
observed. A similar anti-correlation between parallel ion temperature
and equatorial electron density has been reported by
\cite{Schneider:trauger95}. This anti-correlation results from the
increased radiative cooling efficiency of the torus with higher
electron densities.

One of the historical difficulties in understanding the Io plasma
torus has been to separate phenomena that are legitimately time
variable from those that are the result of spatial variations in the
torus that rotate in and out of the observation's field of view. Since
the observed variations have a period very close to the rotation
period of Jupiter, and since the timescales for chemical processes,
such as charge exchange, ionization, recombination, etc., which
produce significant changes in the Io torus are generally much longer
than this \citep{Shemansky88, Barbosa94, Schreieretal98,
  Delamere:bagenal03, Delamereetal04}, the only plausible explanation
is azimuthal variability.

In order to quantify the azimuthal variations in composition, we fit
generalized cosine functions to the ion mixing ratios obtained within
a 50-hour window centered on each data point (e.g. the sinusoidal
curves in Fig.~\ref{mix_sinusoids_vs_time}):

\begin{equation}
  \label{cosine_function}
  M_i(\lambda_{III})=A_i \cos (\lambda_{III}-\phi_i) +c_i
\end{equation}

\noindent where $M_i$ is the ion mixing ratio at the torus ansa; $A_i$
is the amplitude of the compositional variation; $\phi_i$ is the phase
of the variation, i.e. the longitude of the compositional maximum; and
$c_i$ is a constant offset, i.e. the azimuthally-averaged value.  The
amplitude, phase, and constant offset were allowed to vary, while
$\lambda_{III}$, the System III longitude of the ansa (dawn or dusk),
is determined from the time of the observation. Since we are using the
System III longitude of the ansa to determine the phase of the
compositional variation, we are implicitly assuming that the period of
the variations is the 9.925 hour System III Jovian rotation period.
The 50-hour window was chosen as a compromise between the need for
enough data points to produce a robust fit to the data and the need to
minimize the effects of the actual temporal changes in torus
composition evident from Fig.~\ref{p3_mix_vs_time}. The results of the
sinusoidal fits are insensitive to the size of the sliding window used
for windows smaller than $\sim$100 hours. Although there is some
scatter in the data, the simple sinusoidal fits provide a good match
to the conditions in the Io torus.

\subsection{Torus Periodicities}
\subsubsection{Drift of Phase With Time}

\begin{figure}[tbp]
  \includegraphics[scale=.75]{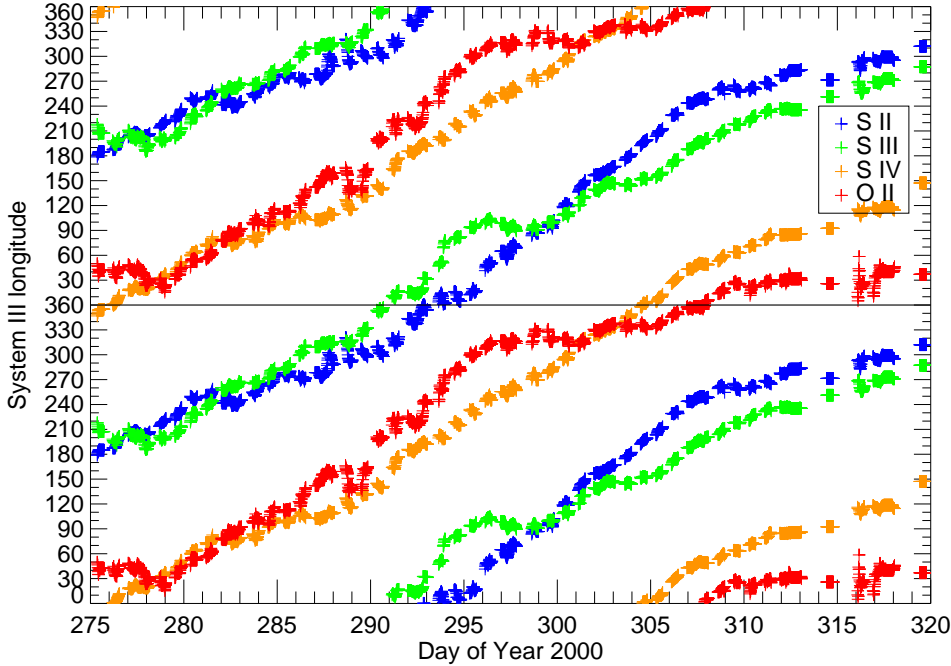}
  \caption[]
  {Phase of azimuthal variations in the composition of the Io torus as
    a function of time. For visual clarity, the top half of the figure
    is a copy of the bottom half. All four ion species show a roughly
    linear trend of increasing phase with time. Data from the dusk
    ansa are shown, results from the dawn ansa are
    similar.\label{dusk_phaseplot}}
\end{figure}

The locations of the peak mixing ratio, i.e. the phase ($\phi_i$) of
the azimuthal variation, for each of the four ion species are plotted
versus time in Fig.~\ref{dusk_phaseplot}. As in
Fig.~\ref{mix_sinusoids_vs_time}, the phase of the azimuthal variation
for \ion{S}{2} tracks with that of \ion{S}{3}, while the phase of
\ion{S}{4} tracks roughly with the phase of \ion{O}{2}, both of which
are shifted approximately 180$^{\circ}$ from the phase of \ion{S}{3}
and \ion{S}{2}. All four of the main ion species in the torus exhibit
phase increases that are roughly linearly with time.  Since the System
III coordinate system is a left-handed coordinate system (System III
longitude increases clockwise when viewed from the north, such that
the sub-observer longitude increases with time for an observer fixed
in inertial space.  See \cite{Dessler83} for more information.) an
increase in phase with time implies that the pattern of compositional
variation is rotating more slowly than Jupiter's magnetic field. The
linear nature of the phase increase implies that the rate of the
subcorotation of the compositional variation is approximately uniform
during the observation period.

\begin{figure}[tbp]
  \includegraphics[scale=.75]{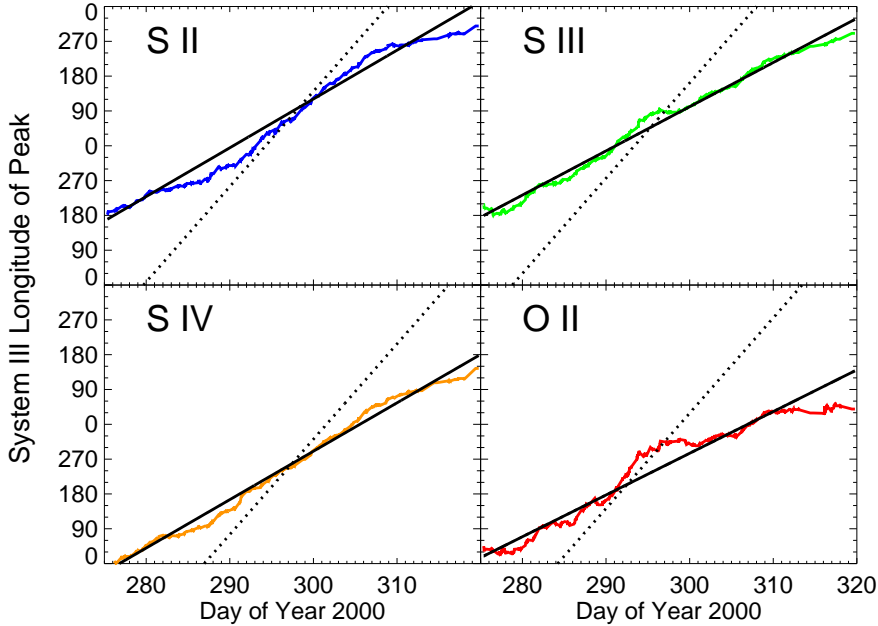}
  \caption[]
  {Linear fit to phase increase with time. The solid curve represents
    the phase of azimuthal variations in composition. The solid line
    is the line best fit to this data. The dotted line represents the
    slope (the intercept of this line is arbitrary) the data would
    have if the plasma in the Io torus were subcorotating at the
    System IV period defined by \cite{Brown95}.
    \label{phase_with_lines}}
\end{figure}

The difference in angular frequency, $\Delta \Omega$, between the
subcorotating compositional variation and the magnetic field of
Jupiter (System III) can be derived from the slope of the phase
increase with time. We used a single linear function to fit the slope
of the phase increases for each of the three sulfur ion species (both
dawn and dusk ansae) simultaneously. The resulting fits are presented
in Fig.~\ref{phase_with_lines}. The dotted lines show the slope that
would be observed if the compositional variations in the torus plasma
were rotating at the System IV period defined by \cite{Brown95}. The
value of $\Delta \Omega$ derived from the three sulfur ion species is
12.5$^{\circ}$/day, corresponding to a period of 10.07 hours---1.5\%
longer than the System III period of 9.925 hours. At a radial distance
of 6~R$_J$, this corresponds to a drift of 1.1 km/s relative to the
magnetic field. The value of $\Delta \Omega$ derived from the UVIS
data is slightly less than half the previously reported values of
$\Delta \Omega \sim$24.3$^{\circ}$/day, which are the basis for the
System IV period \citep{Sandel:dessler88,Woodwardetal94,Brown95}.

Although the phase increase is roughly linear, especially for
\ion{S}{3}, there are several deviations from linearity. For example,
\ion{O}{2} appears to have a greater slope in the period before DOY
295 than after, \ion{S}{2} and \ion{S}{4} show an increase in slope
during the period of DOY 291-309, and all four ion species show a
decrease in slope after day 310.

\subsubsection{Periodograms}

\begin{figure}[tbp]
  \includegraphics[scale=.75]{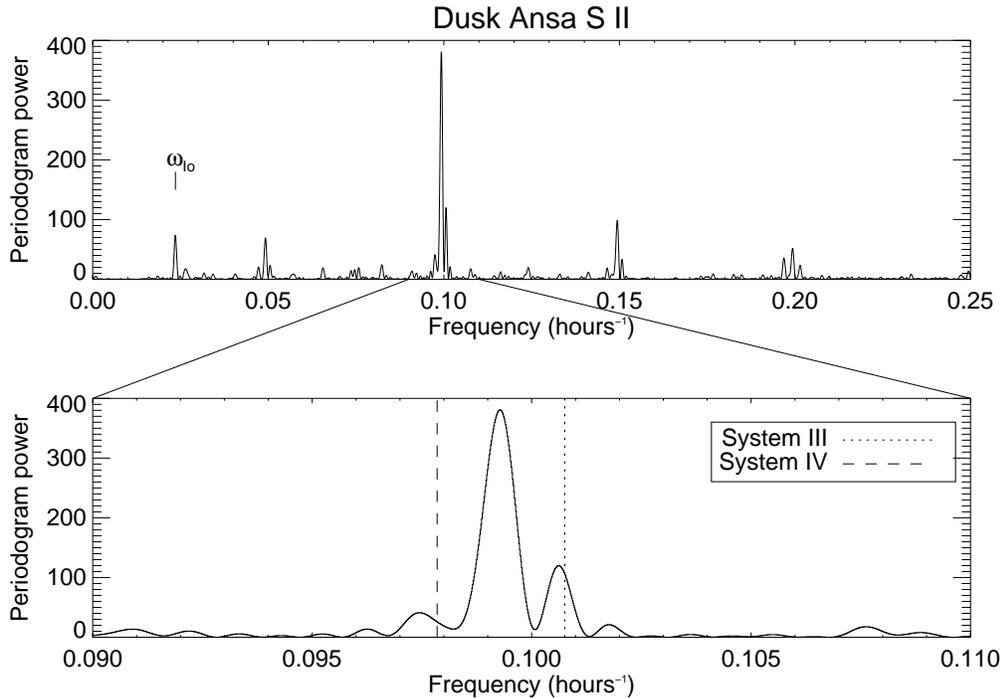}
  \caption[]
  {Lomb-Scargle periodogram of UVIS dusk ansa \ion{S}{2} data. The
    primary peak of the periodogram is found below the System III
    rotation frequency, but above the System IV rotation frequency of
    \cite{Brown95}. \label{s2periodogram}}
\end{figure}

In order to examine our data for periodicities, we have constructed
Lomb-Scargle periodograms \citep{Lomb76,Scargle82,Horne:baliunas86}
using the fast algorithm of \cite{Press:rybicki89}. The periodogram
created from the UVIS dusk ansa \ion{S}{2} data is shown in
Fig.~\ref{s2periodogram}. Periodograms created using from the
\ion{S}{3}, \ion{S}{4}, or \ion{O}{2} data are quite similar;
likewise, periodograms made from the dusk ansa data are virtually
indistinguishable from periodograms made from the dawn ansa data. The
periodogram has a very large peak near a frequency of 0.10 h$^{-1}$,
with smaller peaks occurring near 0.05 h$^{-1}$ and its harmonics. A
secondary peak, located at Io's orbital frequency of 0.024~h$^{-1}$,
is seen only in the \ion{S}{2} data, in contrast to
\cite{Sandel:broadfoot82b} who report a strong correlation of the
brightness of the \ion{S}{3} 685\AA\ feature with Io's phase. No other
significant peaks are present in the UVIS periodograms.

The lower panel of Fig~\ref{s2periodogram} shows the region around the
primary peak in more detail.  The locations of the System III and
System IV frequencies are also shown. The sharp peak seen in the upper
panel actually consists of two closely spaced but separate peaks: the
primary located at a frequency of 0.099277~h$^{-1}$ (period of 10.073
h) and the secondary at 0.10061~h$^{-1}$, slightly below the System
III frequency of 0.10076~h$^{-1}$. The periods obtained from the
frequency of the peak in the periodograms are given in
Table~\ref{periodogram_table}.

\begin{table}[!h]
  \caption{Peak Periodogram Values and Uncertainties}
  \label{periodogram_table}
  \begin{tabular}{lll@{$\, \pm \,$}lll@{$\, \pm \,$}l}
      \ & & \multicolumn{2}{c}{Dusk Ansa} & & \multicolumn{2}{c}{Dawn Ansa} \\
    \hline
    Ion Species & \vline & Period & Uncertainty$^a$ & \vline & Period & Uncertainty$^a$\\
    \hline
    \ion{S}{2} & \vline & 10.073 & 0.0036 h & \vline & 10.073 & 0.0039 h\\
    \ion{S}{3} & \vline & 10.057 & 0.0125  h & \vline & 10.061 & 0.0140  h\\
    \ion{S}{4} & \vline & 10.067 & 0.0039 h & \vline & 10.073 & 0.0039 h\\
    \ion{O}{2} & \vline & 10.049 & 0.0188  h & \vline & 10.051 & 0.0203  h\\
    \hline
    \multicolumn{7}{l}{\footnotesize $^a$ 3$\sigma$ uncertainty derived from
      the 99.7\% value of synthetic datasets}
  \end{tabular}
\end{table}

The probability that the tallest peak in a Lomb-Scargle periodogram is
the result purely of Gaussian-distributed noise in the data (also
known as the false alarm probability) can be derived from the height
of the tallest peak according to the equation:

\begin{equation}
  \label{period_signif}
  F=1-\left[1-\exp^{-h}\right]^{N_i}
\end{equation}

\noindent where $h$ is the height of the tallest peak and $N_i$ is
the number of independent frequencies in the periodogram. It is worth
noting that Eq.~\ref{period_signif} is only valid for the tallest
peak, and cannot be used to assess the significance of any other peaks
present in the periodogram, such as the peak near the System III
frequency.

While it is relatively straightforward to use Eq.~\ref{period_signif}
to obtain the significance of the primary peak in a Lomb-Scargle
periodogram, it is much trickier to obtain an estimate of the
uncertainty in the frequencies of the peaks present, $\Delta$f.
\cite{Kovacs81} derives several expressions for calculating the
$\Delta$f from standard periodogram methods, the derivation assumes
the data contain only a single periodic signal with Gaussian noise,
even data spacing, and no gaps in the data. Although
\cite{Baliunasetal85} found that that these expressions were still
valid in the case of unevenly sampled data, the UVIS data contain
numerous gaps in the data and may also contain signals at multiple
frequencies, rendering this approach invalid.

An order-of-magnitude estimate of $\Delta$f can be made by assuming
that $\Delta$f is equal to the difference in frequency between a
periodic signal that completes $n$ cycles during the observing period
and one that completes $n+\frac{1}{2}$ cycles. The UVIS inbound
staring mode observing period lasted slightly less than 1066 hours,
which leads to a $\Delta$f of $4.69\times10^{-4}$~h$^{-1}$. For a
signal with a period of $\sim$10 hours, this corresponds to an
uncertainty of 0.05 hours.

\begin{figure}[tbp]
  \includegraphics[scale=.75]{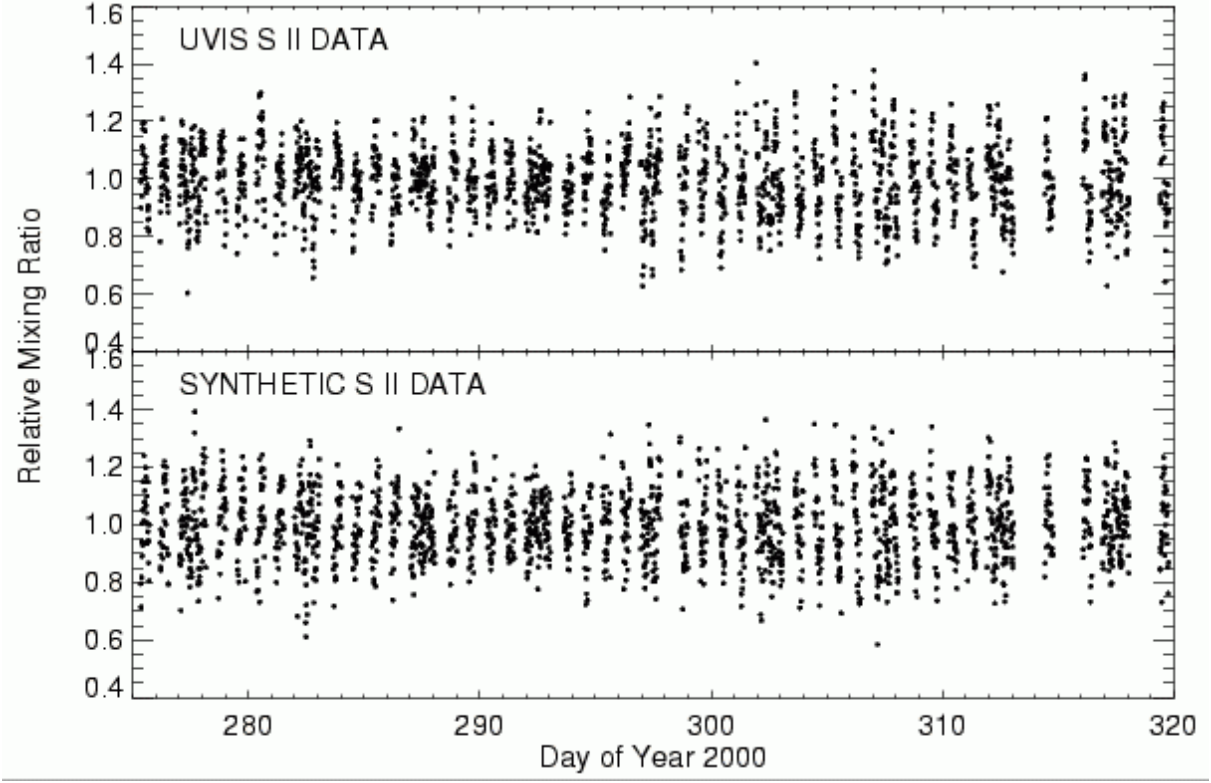}
  \caption[]
  {Comparison of real UVIS \ion{S}{2} data from the dusk ansa with
    synthetic data. The synthetic data consist of a sinusoidal
    variation with a period of 10.073 hours with added Gaussian noise
    sampled at the same times as the UVIS observations. The amplitude
    of the sinusoidal variation in the synthetic data changes with
    time in order to match the real UVIS data.\label{real_vs_synth}}
\end{figure}

This method of estimating $\Delta$f is clearly an oversimplification
as it fails to account for the actual sampling rate or the level of
noise present in the data. We therefore adopt the approach of
\cite{Brown95} and use synthetic data sets to estimate the
uncertainty in our determination of frequency. We constructed 1000
synthetic data sets containing a single periodic signal with a
period of 10.073 hours. This signal had an amplitude similar to that
observed by UVIS and was sampled at the same times as the UVIS
dataset. Gaussian noise, at the level found in the UVIS data, was
also added to the synthetic data. A typical synthetic data set and
the actual UVIS \ion{S}{2} data from the dusk ansa are plotted in
Fig.~\ref{real_vs_synth}.

\begin{figure}[tbp]
  \includegraphics[scale=.75]{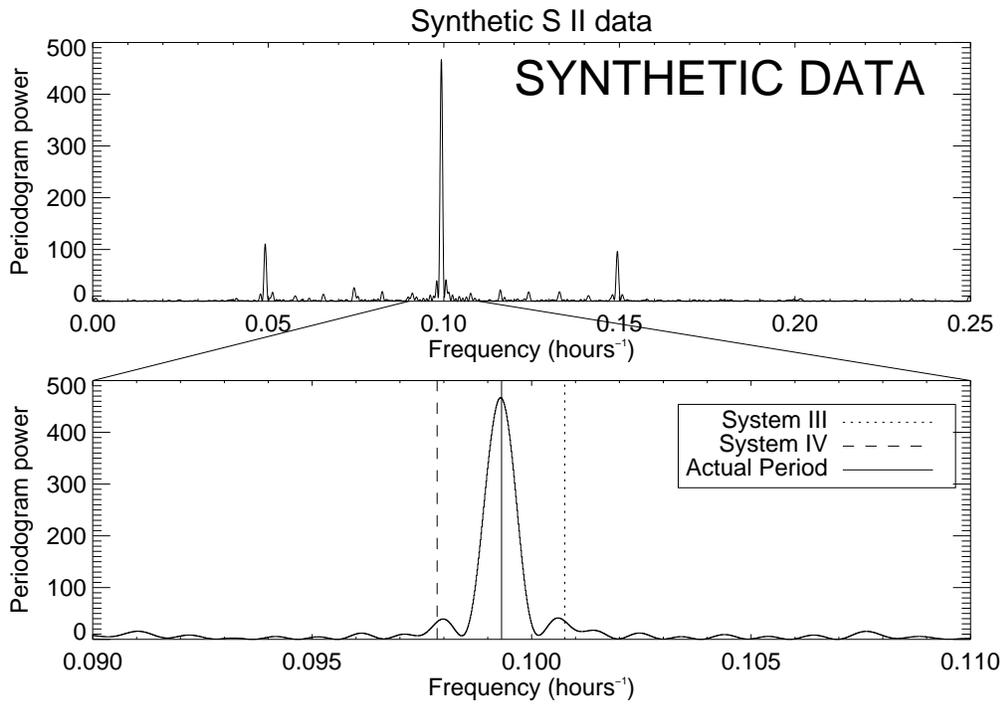}
  \caption[]
  {Periodogram from synthetic \ion{S}{2} data. The primary peak is
    displaced slightly from the actual period in the synthetic data.
    The secondary peaks near 0.05~h$^{-1}$ and 0.15~h$^{-1}$ found in
    Fig.~\ref{s2periodogram} also appear in this periodogram,
    suggesting that they are spurious peaks due to the sampling of the
    UVIS data.
\label{synthperiodogram}}
\end{figure}

We created a Lomb-Scargle periodogram from each synthetic dataset. An
example of a periodogram created from one of the synthetic datasets is
presented in Fig.~\ref{synthperiodogram}. Like the periodogram created
from the real data, c.f. Fig.~\ref{s2periodogram}, the synthetic
periodogram contains a large peak near 0.10~h$^{-1}$ with secondary
peaks near 0.05~h$^{-1}$ and 0.15~h$^{-1}$. The presence of the
secondary peaks in the synthetic periodogram implies that they are the
result of what \cite{Horne:baliunas86} call ``spectral
leakage''---side lobes caused by the data sampling and the finite
observation period. Since there is a 20-hour periodicity in the actual
data sampling, it should not be surprising that spectral power from
the primary peak is aliased to these frequencies.

For each synthetic periodogram, we recorded the difference between the
frequency of the peak and the frequency of the periodic signal
actually present in the synthetic data. We assigned the 68.3\%,
95.5\%, and 99.7\% values a significance of 1$\sigma$, 2$\sigma$ and
3$\sigma$, respectively. This method yielded a 3$\sigma$ estimate of
$\Delta$f for the dusk ansa \ion{S}{2} periodogram of
3.56$\times$10$^{-3}$~h$^{-1}$ or 0.0354\%. The corresponding
3$\sigma$ estimate of the uncertainty in the periods derived from the
peak frequency of the periodograms are given in
Table~\ref{periodogram_table}. Like \cite{Brown95}, we find that our
estimates of $\Delta$f derived from the synthetic data sets are much
smaller than the order-of-magnitude estimate of $\Delta$f derived from
the length of the observation period. However, given that the slope of
the phase increase with time shown in Fig.~\ref{phase_with_lines}
varies over the observing period, it is doubtful whether this result
has any real physical significance. To illustrate this point, we
divided the observing period into three equal parts and made
periodograms from the data in each. The resulting periods derived from
the periodogram peaks are given in
Table~\ref{periodogram_table_3times}. The ion \ion{S}{2} provides an
extreme example, with a period that varies from 9.996--10.137 hours.
This effect can be readily seen in the varying slope of the \ion{S}{2}
curve in Fig.~\ref{phase_with_lines}. Since the value of the period
derived from the location of the periodogram peak depends on both the
time and the duration of the observation window, it should be used
with some caution.

\begin{table}[!h]
  \caption{Peak Periodogram Values of Subdivided Data}
  \label{periodogram_table_3times}
  \begin{tabular}{lllrlr}
    \hline
    \ Ion & Epoch & \vline & Dusk Ansa & \vline & Dawn Ansa \\
    \hline
    \ion{S}{2} & All & \vline & 10.073 & \vline & 10.073 \\
    & Beginning      & \vline & 10.019 & \vline & 10.014 \\
    & Middle         & \vline & 10.137 & \vline & 10.137 \\
    & End            & \vline &  9.996 & \vline &  9.992 \\
    \hline
    \ion{S}{3} & All & \vline & 10.057 & \vline & 10.061 \\
    & Beginning      & \vline & 10.041 & \vline & 10.059 \\
    & Middle         & \vline & 10.032 & \vline & 10.041 \\
    & End            & \vline & 10.023 & \vline & 10.028 \\
    \hline
    \ion{S}{4} & All & \vline & 10.067 & \vline & 10.073 \\
    & Beginning      & \vline & 10.041 & \vline & 10.032 \\
    & Middle         & \vline & 10.087 & \vline & 10.091 \\
    & End            & \vline & 10.014 & \vline & 10.014 \\
    \hline
    \ion{O}{2} & All & \vline & 10.049 & \vline & 10.051 \\
    & Beginning      & \vline & 10.028 & \vline & 10.050 \\
    & Middle         & \vline & 10.014 & \vline & 10.010 \\
    & End            & \vline &  9.983 & \vline &  9.974 \\
  \end{tabular}
\end{table}

\subsection{Amplitude Variations and System III Modulation}
\label{amplitude_variations_subsection}

\begin{figure}[tbp]
  \includegraphics[scale=.75]{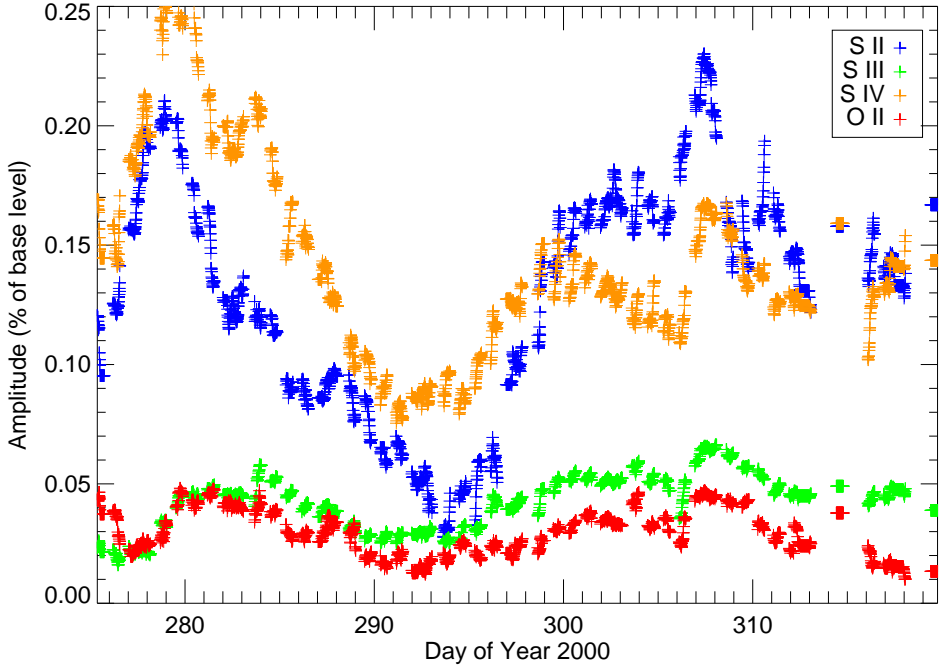}
  \caption[]
  {Relative amplitude (A$_i$/c$_i$) of the azimuthal variations in
    composition as a function of time. The relative amplitudes of the
    major ion species \ion{O}{2} and \ion{S}{3} remain around the few
    percent level, while the relatively minor ion species \ion{S}{2}
    and \ion{S}{4} vary between 4--25\%.\label{dawn_amplitudeplot}}
\end{figure}

Figure~\ref{dawn_amplitudeplot} shows the relative amplitude of the
azimuthal variation in composition ($A_i/c_i$) derived from the
sinusoidal fit to the mixing ratios of the 4 main torus ion species
(cf. Section~\ref{azimuthal_variations_subsection}). All four ion
species have non-zero amplitude for the entire observation period,
suggesting that azimuthal variation in plasma composition is an
omnipresent feature of the Io torus. The amplitudes for the relatively
minor species of \ion{S}{2} and \ion{S}{4}, show dramatic changes with
time.  From the start of the UVIS observations on day 275, the
amplitude of the azimuthal variation in torus composition for both
\ion{S}{2} and \ion{S}{4} increases rapidly with time. When the
amplitudes reach their peak value around day 279, they are nearly a
factor of two greater than when first observed. After reaching their
peak value, both amplitudes fall rapidly to a minimum value around day
293, at which time they are roughly 1/4--1/3 of their peak value. The
amplitudes of the two ion species again increase quickly until day 300
when the \ion{S}{2} amplitude levels out and the amplitude of
\ion{S}{4} decreases somewhat. By day 306, the amplitudes of both ion
species are increasing again, reaching a peak around day 308.

The ion species \ion{O}{2} and \ion{S}{3} also show variations in
amplitude with time, but less dramatically than for \ion{S}{2} and
\ion{S}{4}. The amplitudes for these ion species remain in the range
of 1--6\% during the observing period. Given that \ion{O}{2} and
\ion{S}{3} are the primary ion species for oxygen and sulfur in the Io
torus and that \ion{S}{3} serves as an intermediate product of the
chemical processes that convert \ion{S}{2} into \ion{S}{4} (or vice
versa), this is not surprising. Neither \ion{O}{2} nor \ion{S}{3} show
a well-defined amplitude peak around day 279, although both have
amplitude peaks coincident with the amplitude peaks of \ion{S}{2} and
\ion{S}{4} on day 308.

\begin{figure}[tbp]
  \includegraphics[scale=.75]{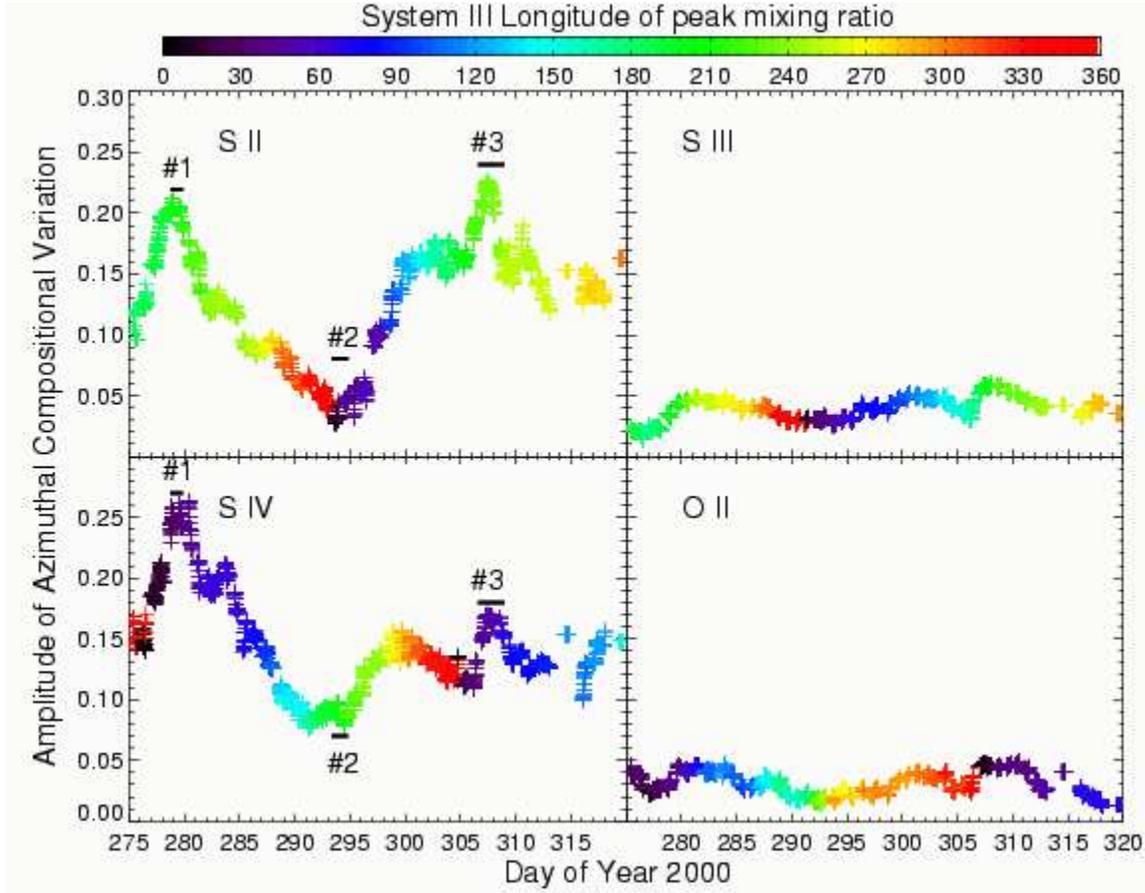}
  \caption[]
  {Relative amplitude of the azimuthal compositional variations as a
    function of time. The color of the plotting symbols represents the
    location (in System III longitude) of the peak mixing ratio. Line
    segments and numbers mark the locations of the three intervals
    used in
    Fig.~\ref{azimuthal_variations_plot}\label{dawn_ampvstime_loncolor}}
\end{figure}

The period of time between the peaks of \ion{S}{2} amplitude is
$\sim$29 days---the same period as the beat between the 9.925-hour
System III rotation period and the observed 10.07-hour periodicity.
This suggests that the amplitude of the azimuthal variation in torus
composition might be modulated by System III. To illustrate this, the
amplitude of the compositional modulation as a function of time is
plotted separately for each species in
Fig.~\ref{dawn_ampvstime_loncolor}.  The color of the plotting symbols
represents the System III phase of the azimuthal variation, i.e. the
location of the mixing ratio peak, for each ion species. The steady
increase of phase with time is readily apparent in all four ion
species. The peaks in amplitude of \ion{S}{2} occur at a phase of
$\lambda_{III} \approx 210^{\circ}$. The \ion{S}{2} amplitude minimum
occurs at a phase of $\lambda_{III} \approx 30^{\circ}$. Conversely,
the amplitude peaks for \ion{S}{4} occur at a phase of $\lambda_{III}
\approx 30^{\circ}$, while the amplitude minimum occurs at a phase of
$\lambda_{III} \approx 210^{\circ}$

\begin{figure}[tbp]
  \includegraphics[scale=.75]{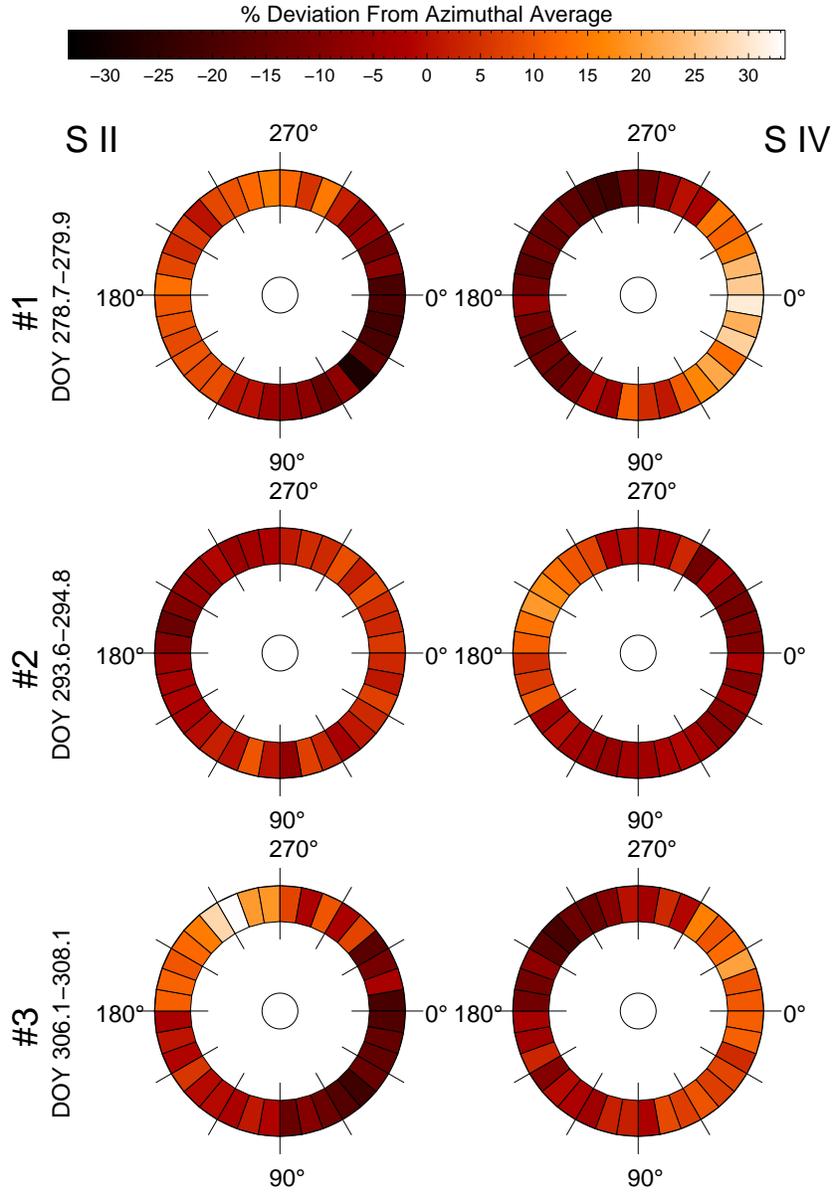}
  \caption[]
  {Graphical representation of the modulation of the amplitude of the
    azimuthal variation with System III longitude. For each time
    interval, the torus has been divided into 36 10$^{\circ}$
    longitude bins. Each bin is colored according to its deviation
    from the average mixing ratio during that time interval. Intervals
    1 and 3 correspond to periods of maximum amplitude, while interval
    2 corresponds to minimum amplitude.  The locations of these
    intervals are marked in Fig.~\ref{dawn_ampvstime_loncolor}.
    \label{azimuthal_variations_plot}}
\end{figure}

Figure~\ref{azimuthal_variations_plot} shows the modulation of the
amplitude of the compositional variation by System III longitude in a
graphical manner for the three times marked in
Fig.~\ref{dawn_ampvstime_loncolor}. In contrast to
Fig.~\ref{dawn_ampvstime_loncolor}, which shows the values of
amplitude and phase derived from the sinusoidal fits,
Fig.~\ref{azimuthal_variations_plot} shows the actual mixing ratio
observed in each 10$^{\circ}$ System III longitude bin, relative to
the azimuthal average. During periods 1 and 3, the mixing ratio of
\ion{S}{2} shows a strong enhancement in the longitude sector
$\lambda_{III}$=180--270$^{\circ}$ and a strong depletion in the
longitude sector $\lambda_{III}$=340--70$^{\circ}$. During the same
periods, the mixing ratio of \ion{S}{4} shows a strong enhancement
between $\lambda_{III}$=330--60$^{\circ}$ and a strong depletion
between $\lambda_{III}$=180--270$^{\circ}$. During period 2,
\ion{S}{2} shows a very weak enhancement between
$\lambda_{III}$=320--50$^{\circ}$ and a weak depletion between
$\lambda_{III}$=160--250$^{\circ}$, while \ion{S}{4} shows an
enhancement between $\lambda_{III}$=160--250$^{\circ}$ and a slight
depletion between $\lambda_{III}$=330--60$^{\circ}$

\section{Discussion}

Initial analysis of the UVIS observations of the Io torus found
long-term azimuthal variations in EUV brightness
\citep{Steffletal04a}, electron density, and electron temperature
(Fig.~\ref{ne_te_vs_sys3}) on the order of $\sim$5\%. However, over
shorter timescales (a few days) the torus is found to exhibit
azimuthal variations in ion composition of up to 25\%. Significant
azimuthal compositional variations were present during the entire
observing period, suggesting that this is the natural state of the Io
torus. Although the primary torus ion species of \ion{S}{3} and
\ion{O}{2} displayed azimuthal variations of only a few percent,
\ion{S}{2} and \ion{S}{4} showed azimuthal variations of up to 25\%.
Models of the torus that treat the mixing ratios of these ion species
as azimuthally uniform must therefore be used with some caution.
Similar caution must be exercised when attempting to apply {\it in
  situ} measurements obtained in one azimuthal region to the torus as
a whole. This may help to explain some of the wide range in electron
densities measured by the Galileo PLS \citep{Frank:paterson01a}.

The azimuthal variations in the composition of the Io torus are
observed to have a period of 10.07 hours---1.5\% longer than the
System III rotation period of Jupiter and 1.3\% shorter than the
System IV period. In ultraviolet, optical, and near-infrared
observations of \ion{S}{2} and \ion{S}{3} obtained between 1979 and
1999, the 10.21 hour System IV period remained remarkably constant
\citep{Roesleretal84,Sandel:dessler88,Brown95,Woodwardetal97,
  Nozawaetal04}, which suggested that this period might be somehow
intrinsic to the Jovian magnetosphere. The presence of a strong
periodicity at 10.07 hours (and corresponding lack of any periodicity
at the System IV period), is therefore rather surprising. While both
\cite{Brown95} and \cite{Woodwardetal97} observed abrupt changes in
the phase of the azimuthal variation in brightness of [\ion{S}{2}]
6731\AA\ which caused their initial analysis to identify a spurious
periodicity of 10.16 hours, it is evident from
Fig.~\ref{dusk_phaseplot} that no such change in phase occurred during
the UVIS observations.

Given the phenomenological similarity between the UVIS 10.07 hour
periodicity and the System IV periodicity, we propose that the same
physical mechanism is responsible for both. It is plausible that the
factor of 3--4 increase in the amount of neutrals supplied to the
torus in September 2000 \citep{Delamereetal04} altered the mechanism
responsible for producing the System IV period in such a manner that a
10.07 hour period was produced. Based on measurements of Iogenic dust
by the Galileo Dust Detector System \citep{Kruegeretal03}, such events
occur relatively infrequently (only one event of this magnitude was
detected during 6.5 years of observations). If future observations of
the Io torus detect periodicity at the 10.21 hour System IV period
(and not at 10.07 hours), it would suggest that the intermediate
period observed by UVIS was a result of the neutral source ``event''
that occurred in September 2000. Ground-based observations in December
2000 (roughly one month after the end of the UVIS staring-mode
observations presented in this paper) found the brightness of torus
[\ion{S}{2}] 6712\AA\ and 6731\AA\ emissions varied with a period of
10.14$\pm$0.11 hours \citep{Nozawaetal04}, which suggests the torus
periodicity might have been returning to the ``typical'' System IV
period. However, given the relatively large uncertainty in this value,
it is consistent with both the 10.07 hour period measured by UVIS and
the canonical 10.21 hour System IV period, so no firm conclusions can
be drawn.

It is important to reiterate that we do not (and can not) directly
measure the rotation speed of the torus plasma with UVIS. Rather, we
derive the rotation period of azimuthal variations in the composition
of the Io torus. This value will be affected by both the actual
rotation speed of the plasma and any spatial/temporal changes in the plasma
composition resulting from torus chemistry.

If the 10.07 hour periodicity in the UVIS data were produced directly
by the subcorotation of torus plasma, the plasma, when averaged over
the UVIS field of view and weighted by its EUV emissions, would need
to lag the local corotation velocity by 1.5\% ($\sim$0.19 km/s/R$_J$).
In order to maximize signal-to-noise in the torus spectra, spectra
from three rows on the detector were averaged together, as described
in Section~\ref{observation_section}. If, however, spectra are
extracted from just a single row on the detector, the same 10.07 hour
periodicity is evident throughout the observing period, despite the
factor of 3 decrease in the size of the field of view. Furthermore,
this holds true for both the dawn and the dusk ansae of the torus
which look at slightly different ranges of radial distance. Given
these considerations and the changes in viewing geometry of the UVIS
observations over the 45-day observing period, the only feasible way
to produce such a periodicity directly from subcorotating plasma is if
the deviation from corotation were constant with radius. This,
however, is not supported by observations of the radial velocity of
\ion{S}{2} that show a strong variation in the subcorotation speed of
the plasma with radial distance \citep{Brown94b,Thomasetal01}.  The Io
torus must have been in a radically different state during the Cassini
encounter than it was in 1992 (when the observations of
\cite{Brown94b} were made) and 1999 (when the observations of
\cite{Thomasetal01} were made) if the subcorotation of the torus
plasma is directly responsible for the periodicity in the UVIS data.
Since there are also theoretical arguments against producing such
periodicities in the torus directly from plasma subcorotation, i.e.
that energy must be supplied continuously to the torus in just the
right place and amount to keep the pattern coherent despite the
changes in radial distance \citep{Dessler85,Sandel:dessler88}, we
consider this possibility unlikely.

Instead, the UVIS observations are entirely consistent with the theory
proposed by \cite{Brown94a} that the System IV periodicity is the
result of the pattern speed of a compositional wave propagating
through the Io torus. While the individual particles of the torus lag
corotation by an amount appropriate for their radial distance, the
group velocity of the compositional wave lags rigid corotation by
1.5\%.

The amplitude of the azimuthal variation in composition appears to be
modulated by its position relative to System III longitude.  During
times when the peak (minimum) in \ion{S}{2} (\ion{S}{4}) mixing ratio
is aligned with a System III longitude of $210^{\circ} \pm 15^{\circ}$
the amplitude of the azimuthal variation in composition is enhanced.
When the peak (minimum) in \ion{S}{2} (\ion{S}{4}) mixing ratio is
aligned with a System III longitude of $30^{\circ} \pm 15^{\circ}$,
the amplitude of the variation is diminished, i.e. the torus becomes
more azimuthally uniform. Since UVIS observed only 1~1/2 modulation
cycles, it is difficult to say whether the apparent modulation by
System III longitude is real or just coincidental. However, similar
modulations in the brightness of the \ion{S}{3} 685\AA\ feature
observed by the Voyager 2 UVS, the probability of detecting nKOM
emission with the Voyager PRA instruments, and the brightness of torus
[\ion{S}{2}] 6731\AA\ emissions were reported by
\cite{Sandel:dessler88} and \cite{Schneider:trauger95} and may also be
present in the data of \cite{Pilcher:morgan80}.

In light of the UVIS observations of a subcorotating azimuthal
variation in composition whose amplitude is modulated by its position
relative to System III longitude, several apparently contradictory
observations of the Io torus can be explained. First, because the
azimuthal variation subcorotates relative to System III, the phase of
the variation (i.e.  the location of peak) should be observed over the
full 360$^\circ$ range of longitude. However, because the amplitude of
the azimuthal variation is greatest when the peak in \ion{S}{2} mixing
ratio is located near $\lambda_{III}=200^{\circ}$ azimuthal variations
in the brightness of \ion{S}{2} emissions will be preferentially
detected in the ``active sector'' centered around
$\lambda_{III}=200^{\circ}$.

The detection of azimuthal variability in the brightness of the
[\ion{S}{2}] 6716\AA/6731\AA\ and 4069\AA/4076\AA\ doublets but not in
the brightness of the [\ion{O}{2}] 3726\AA/3729\AA\ doublet
\citep{Morgan85}, arises from the fact that the amplitude of the
azimuthal variation of the \ion{S}{2} mixing ratio ranges from
10--25\% while the amplitude of the azimuthal variation of the
\ion{O}{2} mixing ratio is $\leq$5\%. The correlation of \ion{S}{2}
brightness with \ion{S}{3} brightness observed by \cite{Raueretal93}
is also consistent with the correlation of the \ion{S}{2} and
\ion{S}{3} mixing ratios observed by UVIS.

The two week transition between azimuthally varying and azimuthally
uniform states observed by \cite{Pilcher:morgan80} is just the
manifestation of the modulation period, assuming that the modulation
period was $\sim$14 days (which corresponds to the beat between the
9.925 hour System III period and the 10.214 hour System IV period).
The $\sim$14 day modulation period of the amplitude of the azimuthal
compositional variation also explains why \cite{Morgan85} observed a
strong azimuthal variation in the brightness of the torus
[\ion{S}{2}] 6731\AA\ emission with a peak near
$\lambda_{III}=180^{\circ}$ during the period of 14--17 February
1981, while \cite{Brown:shemansky82} detected no significant
azimuthal variation in the same emission line on 23--24 February
1981. Furthermore, the subcorotation and amplitude modulation of the
azimuthal variation in composition explains why the Voyager 2 UVS
observed only a weak azimuthal variation in the brightness of the
\ion{S}{3} 685\AA\ feature with a peak between $330^{\circ} <
\lambda_{III} < 40^{\circ}$ on day 122 of 1979 (the amplitude of the
azimuthal variation was at a minimum), a stronger azimuthal
variation (the azimuthal variation had reached it's minimum
amplitude with the \ion{S}{3} peak near 20$^\circ$ and was
increasing in amplitude) with a peak between $40^{\circ}
<\lambda_{III} < 100^{\circ}$ on day 124 of 1979 (assuming the
azimuthal pattern had the 10.2 hour System IV period, the phase
should increase by $\sim$24$^{\circ}$/day), and no significant
System III variation when the spectra were averaged over the 44 day
pre-encounter period (the System III longitude of the peak varies
with time). Whereas the failure of \cite{Gladstone:hall98} to detect
any significant variation in the brightness of torus emissions with
System III longitude results from the averaging of EUVE data
obtained during the interval from 19--24 June 1996. During this
time, the peak of the azimuthal variation would have shifted by over
120$^\circ$ in System III longitude.

Finally, it is also interesting to note that the two largest torus EUV
luminosity events reported by \cite{Steffletal04a} occurred on days
280 and 307, near the modulation peaks. During these events, which
last for roughly 20 hours, the power radiated by the Io torus in the
EUV increases rapidly by $\sim$ 20\% before gradually returning to the
pre-event level. Since several other brightening events occurred
throughout the 45-day observing period, the timing of the two largest
events may be coincidental.

The next step is to model the UVIS observations by extending the torus
chemistry model of \cite{Delamereetal04}. Preliminary results suggest
that the interaction of a subcorotating (at 10.07 hours),
azimuthally-varying source of hot ($\sim$55 eV) electrons with a
corotating (i.e. fixed in System III), azimuthally-varying source of
hot electrons can produce torus behavior that is both qualitatively
and quantitatively similar to the UVIS observations. The 28.8 day
modulation period arises naturally from the beating of the 10.07 hour
period with the 9.925 hour System III period. While it is not too
difficult to imagine a source mechanism capable of producing hot
electrons in amounts that vary as a function of System III longitude,
it is far from obvious what could cause an additional,
azimuthally-varying pattern of hot electrons to rotate with a period
1--3\% slower than the System III period. At present we know of no
suitable physical mechanism capable of producing such behavior.

\section{Conclusions}

We have presented an analysis of the temporal and azimuthal
variability of the Io plasma torus during the Cassini encounter with
Jupiter. Our main conclusions are:

\begin{enumerate}
\item The torus exhibited significant long-term compositional changes
  during the UVIS inbound observing period. These compositional
  changes are consistent with models predicting a factor of 3--4
  increase in the amount of neutral material supplied to the torus in
  early September, 2000. These results are discussed in more detail by
  \cite{Delamereetal04}.

\item Persistent azimuthal variability in torus ion mixing ratios,
  electron temperature, and equatorial electron column density was
  observed. The azimuthal variations in \ion{S}{2}, \ion{S}{3}, and
  electron column density mixing ratios are all approximately in phase
  with each other. The mixing ratios of \ion{S}{4} and \ion{O}{2} and
  the torus equatorial electron temperature are also approximately in
  phase with each other, and as a group, are approximately
  180$^{\circ}$ out of phase with the variations of \ion{S}{2},
  \ion{S}{3}, and equatorial electron column density.

\item The phase of the observed azimuthal variation in torus
  composition drifts 12.2$^{\circ}$/day, relative to System III
  longitude. This implies a period of 10.07 hours, 1.5\% longer than
  the System III rotation period. This period is confirmed by
  Lomb-Scargle periodogram analysis of the UVIS data.

\item The relative amplitude of the azimuthal variation in composition
  is greater for \ion{S}{2} and \ion{S}{4}. These species have
  relative amplitudes that vary between 5--25\% over the observing
  period. The major ion species, \ion{S}{3} and \ion{O}{2}, have
  relative amplitudes that remain in the range of 2--5\%.

\item The amplitude of the azimuthal compositional variation appears
  to be modulated by its position relative to System III longitude
  such that when the peak in \ion{S}{2} mixing ratio is aligned with a
  System III longitude of 210$\pm$15$^{\circ}$ the amplitude is
  enhanced, and when the peak in \ion{S}{2} mixing ratio is aligned
  with a System III longitude of 30$\pm$15$^{\circ}$ the amplitude is
  diminished.

\end{enumerate}

\acknowledgements
\noindent {\bfseries Acknowledgements}\\
Analysis of the Cassini UVIS data is supported under contract
JPL~961196. FB acknowledges support as Galileo IDS under contract
JPL~959550. The authors wish to thank Ian Stewart, Bill McClintock,
and the rest of the UVIS science and operations team for their
support.


\end{document}